\documentclass[amsmath,amssymb,amsfonts,aps,prl,twocolumn,superscriptaddress,footinbib]{revtex4-2}

\usepackage{graphicx}
\usepackage{dcolumn}
\usepackage{bm}
\usepackage[normalem]{ulem}
\usepackage{xcolor}
\usepackage{comment}
\usepackage{amsmath}
\usepackage{siunitx}
\usepackage{dsfont}
\usepackage{float}
\usepackage{placeins}
\usepackage[shortlabels]{enumitem}

\newcommand{\up}{\uparrow}
\newcommand{\dn}{\downarrow}
\DeclareSIUnit\angstrom{\text{Å}}

\begin{document}

\title{Beyond spin models in orbitally-degenerate open-shell nanographenes}

\author{J. C. G. Henriques}
\affiliation{International Iberian Nanotechnology Laboratory (INL), Av. Mestre Jos\'e Veiga, 4715-330 Braga, Portugal}
\affiliation{Universidade de Santiago de Compostela, 15782 Santiago de Compostela, Spain}

\author{D. Jacob}
\affiliation{Departamento de Pol\'{i}meros y Materiales Avanzados: F\'{i}sica, Qu\'{i}mica y Tecnolog\'{i}a, Universidad del Pa\'{i}s Vasco UPV/EHU,
  Av. Tolosa 72, E-20018 San Sebasti\'{a}n, Spain}
\affiliation{IKERBASQUE, Basque Foundation for Science, Plaza Euskadi 5, E-48009 Bilbao, Spain}

\author{A. Molina-Sánchez}
\affiliation{Institute of Materials Science (ICMUV), University of Valencia, Catedr\'{a}tico Beltr\'{a}n 2, E-46980 Valencia, Spain}

\author{G. Catarina}
\affiliation{nanotech@surfaces Laboratory, Empa---Swiss Federal Laboratories for Materials Science and Technology, 8600 D\"{u}bendorf, Switzerland}

\author{A. T. Costa}
\affiliation{International Iberian Nanotechnology Laboratory (INL), Av. Mestre Jos\'e Veiga, 4715-330 Braga, Portugal}

\author{J. Fern\'andez-Rossier}
\altaffiliation[On permanent leave from ]{Departamento de F\'isica Aplicada, Universidad de Alicante, 03690 San Vicente del Raspeig, Spain.}
\affiliation{International Iberian Nanotechnology Laboratory (INL), Av. Mestre Jos\'e Veiga, 4715-330 Braga, Portugal}

\date{\today}

\begin{abstract} 
The study of open-shell nanographenes has relied on a paradigm where spins are the only  low-energy degrees of freedom. Here we show that some nanographenes  can host  low-energy excitations that include strongly coupled spin and orbital degrees of freedom. The key ingredient is the existence of orbital degeneracy, as a consequence of leaving the benzenoid/half-filling  scenario. We analyze  the case of nitrogen-doped triangulenes, using both density-functional theory and Hubbard  model multiconfigurational and random-phase approximation calculations. We find a rich interplay between orbital and spin degrees of freedom that confirms the need to go beyond the spin-only paradigm,   opening a new venue in this field of research.
\end{abstract}
\maketitle

The prediction that  graphene fragments, graphene quantum dots or graphene islands, could have an open shell ground state, with finite electronic spin $S$, goes back many decades~\cite{clar53,ovchinnikov78}.  Their experimental study has been hampered by their large chemical reactivity,  so that only ensemble measurements with paramagnetic resonance could be used. With the advent of on-surface synthesis~\cite{cai2010}, combined with surface scanning probes, the study of the magnetic properties of these fascinating systems has undergone a revolution~\cite{ruffieux16,pavlivcek2017,mishra2019b,turco23,sanchez}. It is now possible to study supramolecular structures where open-shell nanographenes, such as triangulenes, assemble to form dimers~\cite{mishra2020,krane23}, rings~\cite{mishra21,hieulle2021},  chains~\cite{mishra21}, and  two-dimensional lattices~\cite{delgado23}. The magnetic properties of these structures can be accounted for by spin models~\cite{mishra21,Jacob22,catarina2023,du2023orbital} that feature exotic properties such as fractionalization and symmetry protected topological order~\cite{mishra21}. 

Here we enhance this paradigm to address the case of molecules whose ground state has a degeneracy larger than $2S+1$ on account of their orbital degeneracy. This is motivated in part  by the recent synthesis of nitrogen doped triangulenes~\cite{wang2022,lawrence23,calupitan2023emergence,vegliante2023surface,wei2022solution} for which  orbital degeneracy can be expected, as we discuss below. This situation has long been studied in the context of transition metal oxides~\cite{kugel82,tokura00}, and it is known to bring new electronic phenomena, such as quantum melting of magnetic order~\cite{feiner97}, solitonic phases~\cite{brey05}, and spin-orbital entanglement~\cite{lundgren12}. The observation of orbital Kondo effect  in carbon nanotubes \cite{jarillo05}, and proposal to use coupled spin-orbital qubits in that system \cite{rohling12}, provide additional motivation for the present work.

\begin{figure}
    \centering
    \begin{tabular}{cc}
        \includegraphics{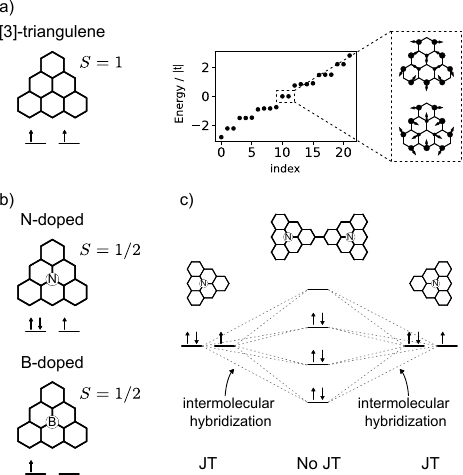}
    \end{tabular}
    \caption{a) Schematic representation of a [3]-triangulene and its single particle spectrum obtained from a tight binding model with first neighbor hopping $t$, and third neighbor hopping $t_3 = t/10$. In the site representation of the two zero mode wave functions, the circle size refers to the absolute value of the wave function, and the arrows encode the phase. b) Example of two functionalized [3]-triangulenes where the central carbon atom is replaced by either nitrogen or boron.  c) Pictorial representation of intermolecular hybridization which lifts the degeneracy found for the zero modes of the monomers. Only the monomers show Jahn-Teller (JT) distortion.  
    }
    \label{fig:1}
\end{figure}

In Fig. \ref{fig:1}a we show a regular [3]-triangulene, together with its single particle spectrum, that features two $C_3$ symmetric degenerate levels at zero energy. Depicted in Fig. \ref{fig:1}b
are examples of molecules that, ignoring Jahn-Teller (JT) distortion~\cite{JahnTeller37}, host a ground state with both orbital and spin degeneracy. This prediction is based on the existence of an odd number of electrons, that ensures a minimum spin degeneracy of 2, corresponding to $S_z=1/2$,  and the two-fold orbital degeneracy of the highest occupied molecular orbital, predicted by tight-binding calculations.  In the cases considered here, two main ingredients are at play. First,  $C_3$ symmetry leads to vanishing of the zero mode wave functions at the central atoms and thus prevents the lifting of the degeneracy of the orbital doublets. Second, molecules with an  odd number of electrons, obtained, for example, via functionalization of the molecule through the substitution of a carbon by a boron or a nitrogen atom (an alternative, not addressed here, is the presence of non-benzenoid rings).

Importantly, in  both cases the Ovchinnikov-Lieb rule\cite{ovchinnikov78,Lieb1989,fernandez07} cannot be applied to these systems.  Lieb theorem applies for the Hubbard model in bipartite lattices at half filling. In functionalized benzenoid molecules, the addition/removal of an  electron takes the system away from half-filling (and for non-benzenoid molecules, the lattice is no longer bipartite). Hence,  the spin of the ground state of these  molecules cannot be easily anticipated. 

Exactly as in the case of transition metal oxides, JT  distortions do occur when a single molecule is considered. In dimers, however, the orbital degeneracy is slightly lifted by intermolecular hybridization (see Fig. \ref{fig:1}c)  protecting the symmetry of the molecules from JT distortions, and preserving the availability of extra orbital states that entangle with the spin degrees of freedom. 

The central question that we address in this work is the following: what are the low-energy properties of supramolecular structures made with the orbitally degenerate $S=1/2$ building blocks. In the following we focus on the case of nitrogen doped triangulenes, also known as Aza-[3]-triangulenes (A3T), although we expect our main results will apply to other systems. We consider A3T as the building block of two classes of structures considered in the literature for related molecules that are either closed shell or open-shell without orbital degeneracy:    dimers\cite{mishra2020,Jacob22,krane23} and  honeycomb crystals\cite{jing18,zhou20,sethi2021,ortiz22,delgado23,catarina2023}. We model both with density functional based calculations and with a Hubbard model, treated at three levels of approximation: mean field (both dimers and crystals), random phase approximation (for the crystal) and multi-configurational methods (for the dimer).  
\begin{figure}
    \centering
    \includegraphics{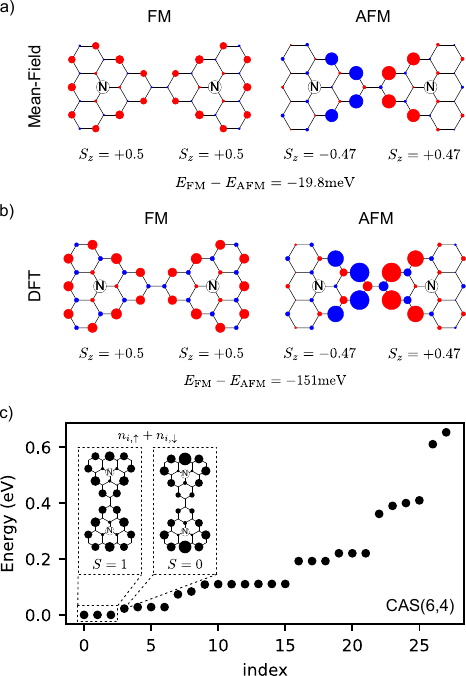}
    \caption{MFH, DFT and CAS results for A3T-dimer.
    (a,b) Spin densities of FM and AFM solutions for the ground state (GS)
    computed with (a) MFH using $U = |t|$, $V_0 = -4$eV, $V_1 = -0.85$eV, $t_3 = t/10$, $t_2 = 0$ and $t = -2.7$eV,
    and (b) DFT using PBE together with 6-311G basis set. Also depicted are the values of $S_z$ per triangulene for each solution and the energy difference between FM and AFM
    solutions. The radius of each circle is proportional to the atomic spin, and red (blue) denote spin-up (spin-down). 
    c) Energies obtained from a CAS(6,4) calculation. The inset shows the charge at each site ($n_{i,\up} + n_{i,\dn})$ for the ground state triplet and the first singlet. The same parameters of panel a) were considered.
    \label{fig:fig2}
    }
\end{figure}

We first discuss the case of the A3T dimer. We have performed mean-field Hubbard (MFH) model as well as density functional theory (DFT) calculations. For the latter we have employed the Gaussian16 quantum chemistry package~\cite{g16} with the PBE functional~\cite{PBE,SM}. For the Hubbard model we use the same parameters as in previous work for triangulenes, with first neighbour hopping $t=-2.7$eV, third neighbour hopping $t_3= t/10$ and $U = |t|$ \cite{Jacob22}; furthermore, to account for the doping an additional on-site potential on the nitrogen site, $V_0 = -4$eV, and on its three nearest neighbors,  $V_1= -0.85$eV, are included in the model. The values of $V_0$ and $V_1$ are close to those reported in the literature for nitrogen doped-graphene \cite{tison2015electronic}, but we choose $V_1$ as to make the bands of the non-magnetic phase of the two dimensional crystal similar to those calculated with DFT~\cite{SM}.

We consider  ferromagnetic (FM)  and antiferromagnetic (AFM) solutions, both in MFH and in DFT. Both methods predict the FM configuration to have lower energy than the AFM solution, and give similar magnetic maps for the FM and AFM arrangements, with the same total value of $S_z$ per triangulene, as shown in Fig. \ref{fig:fig2}. We note that a ferromagentic ground state was found by Yu and Heine~\cite{yu23dimer} using  using the PBE0 hybrid functional. Importantly, both DFT and mean-field Hubbard approaches show a feature that cannot be captured by a simple spin model:  the difference between the magnetization map  of the FM and AFM solutions goes beyond a mere change of sign in the magnetization of one A3T (see Fig.\ref{fig:fig2}a,b), at odds with the case of conventional open-shell dimers~\cite{mishra2020} and with existing literature of magnetic nanographenes~\cite{munoz09,yazyev10,zhang10}.
Furthermore, we note that the energy difference between FM and AFM states differs significantly in the two approaches:  $\sim$19~meV for MFH and $\sim$151~meV for DFT.  In contrast,
the discrepancy across methods is much smaller for the two-dimensional crystal (see below). We attribute this to the fact that the single-particle parameter $V_1$ has been chosen as to obtain a good agreement with the single-particle states for the crystal, and not for the dimer. Moreover, since $V_1$ is the electrostatic potential created by Nitrogen on its first  neighbouring sites, which  in turn depends on the screening, one may expect $V_1$ to be different for a molecule and  a crystal. Here, for simplicity, we consider a single value.

In order to verify that the  spin-orbital  interplay  is not a shortcoming of the mean-field approximation, but rather a feature of the molecules, we carry out a multi-configurational calculation, in a restricted Hilbert space, using the complete active space (CAS) approach~\cite{ortiz19,Jacob22,catarina2023,henriques2023}. 
We first solve the single particle problem, and restrict the many-body Hilbert space to the four single-particle orbitals closest to zero energy, over which six electrons are distributed (similar to what is depicted in Fig. \ref{fig:1}c). When implemented with a pair of orbitally non-degenerate  open-shell molecules with spin $S_1$ and $S_2$, the CAS method yields spectra  with a manifold of low-energy states of dimension $(2 S_1+ 1)(2 S_2 +1)$, well separated from the next excitations \cite{Jacob22,catarina2023,henriques2023}. 
This permits one to model the fermionic low-energy states in terms of a spin Hamiltonian.  Thus, since the A3T monomer has $S=1/2$, the naive expectation for the CAS low-energy spectrum would be a manifold with four states,  corresponding to a triplet and a singlet.  In contrast, our calculations, depicted in Fig. \ref{fig:fig2}c, show a spectrum with 16 states, that includes  4 singlets and 4 triplets, followed close in energy by an additional set of 12 states (corresponding to ionic configurations). Therefore, the orbital degeneracy of the monomers, that is quenched at the single-particle level due to intermolecular hybridization, re-emerges in the interacting limit. The magnetization map of the $S_z = +1$ component of the ground state manifold \cite{SM} is  in excellent agreement with those obtained with  MFH and DFT. In the inset of Fig. \ref{fig:fig2}c   the charge maps for the ground state ($S=1$) and the lowest energy excited state ($S=0$)  are shown to be different.  This implies a spin-dependent  occupation of the underlying molecular orbitals, demonstrating  the interplay between spin and orbital degrees of freedom. Further evidence of this is discussed in \cite{SM}. At last, we note that the lack of a clear energy separation between the manifold of the first 16 energy states, and the 12 higher energy ones, raises the question of whether charge fluctuations will also play a role, invalidating the use of models such as the one proposed by Kugel and Khomskii \cite{kugel82}. This will be the subject of a future study.

\begin{figure}
    \centering
    \includegraphics[width=0.95\linewidth]{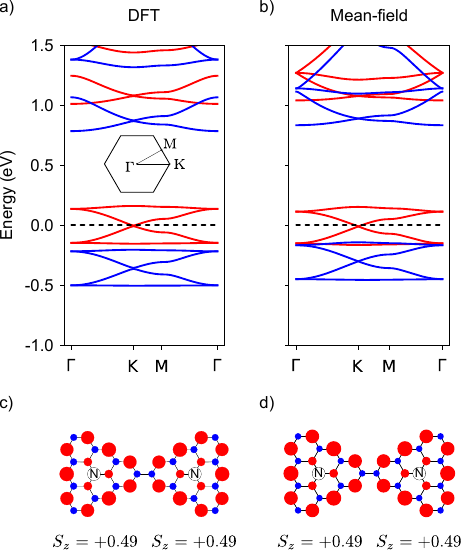}
    \caption{ 
Energy bands and magnetic moment distributions for A3T for the ferromagnetic configuration obtained with DFT (panels a and c) and MFH model (panels b and d). For the MFH model calculation we adopted the same parameters as in Fig. \ref{fig:fig2}.}
    \label{fig:3}
\end{figure}

We now consider the electronic properties of the two-dimensional honeycomb lattices whose unit cells are the dimers considered in Fig. \ref{fig:fig2}. As we did in the case of molecular dimers, we compute their electronic energy bands using two approaches, MFH model and DFT based calculations. The DFT calculations have been performed at the local-density approximation (LDA) using PBE functionals \cite{Hamann2013,van_setten_pseudodojo_2018,SM}. We have optimized the structure in the ferromagnetic phase, that is the ground state, and the calculations of the band structure of the antiferromagnetic and non-magnetic phase are performed on top of
these optimized atomic positions. We find an excellent agreement between DFT and MFH: both methods predict a half-metallic FM ground state with the Dirac crossing at $K$, with very similar bands, isomorphic to the $p_x-p_y$ model \cite{wu2007flat}, and a very similar magnetization map with a total spin $S_z \approx 1/2$ per triangulene (see Fig. \ref{fig:3}). From the DFT calculation it is found that the FM solution lies $\approx 98$ meV below in energy when compared to the AFM one, while an energy difference of $\approx 67$ meV is found in the MFH approach. This agreement validates our choice of parameters for the Hubbard model.

The origin of the peculiar energy bands can be understood as follows. For the non magnetic (NM) case \cite{SM}, where both spin channels are degenerate,  the Fermi surface lies at the point where a dispersive band touches a flat band. Therefore, the Fermi surface is a point at $\Gamma$ giving rise to a so-called singular flat band predicted to have non-trivial quantum geometric properties \cite{rhim20}. In the FM phase (Fig. \ref{fig:3}) the spin-up and spin-down bands approximately keep the line-shape of the NM solution and are vertically shifted in energy in opposite directions. To preserve the number of electrons in the system, one spin channel sees its four low-energy bands completely filled, while for the other one only two bands can be completely occupied, thus pushing the Fermi level to the Dirac point, and giving rise to the half-metal character of the FM solution.

We have also computed AFM solutions for the A3T crystals \cite{SM}. Similar to the FM case, the AFM solution is also a half-metal, but this time with the Fermi-level crossing the bands at a single point in the center of the Brillouin zone. As in the case of molecular dimers, the difference between AFM and FM magnetization goes beyond flipping-over the magnetic moments in one of the triangulenes, although the difference is smaller than in the dimer \cite{SM}.

The fact that the magnetic phases are predicted to be conducting already provides a strong hint that their spin properties cannot be described with a spin model. However, it is often the case that the spin dynamics of conducting ferromagnets are described with spin models\cite{liechtenstein84,costa03}. For the FM solution with $S=1/2$ per A3T molecule, the Heisenberg model in the honeycomb lattice would be an obvious candidate.  In order to test this possibility, we calculate collective modes of the FM phase using the Random Phase Approximation (RPA) \cite{Barbosa2001,peres2004,catarina2023}.
This method computes the spin susceptibility matrix in the energy-momentum ($\omega,\vec{k}$) plane. The poles of this matrix correspond to spin excitations of the system. In this approach there are two broad classes of excitations: magnon modes with a well defined energy versus momentum dispersion, such as the Goldstone modes associated to a broken-symmetry ground state, and electron-hole type of excitations that define a continuum in the $(\omega,\vec{k})$ plane. 

\begin{figure}
    \centering
    \includegraphics[width=\columnwidth]{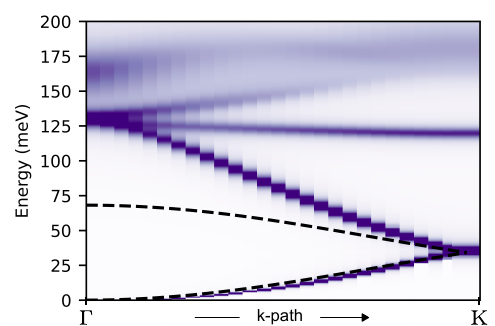}
    \caption{Density plot of the magnon spectral density ($-\mathrm{Im}\chi^{+-}$) in the wave vector-energy plane along the $\Gamma-K$ line. Dispersion relations for spin-flip excitations can be inferred from the dark spots associated with the maxima of the spectral density. The black dashed lines are a fit to the magnon dispersion relation of a nearest-neighbor FM Heisenberg model in a honeycomb lattice.
    }
    \label{fig:magnondispersions}
\end{figure}

Our RPA calculations for the collective spin excitations of the FM ground state, depicted in Fig. \ref{fig:magnondispersions}, show  at least four well defined magnon branches. This is in stark contrast with  both  the  FM Heisenberg spin model in the honeycomb lattice and with analogous RPA results for un-doped triangulene crystals\cite{catarina2023}, for which we find only {\em two} branches of spin excitations. The departure of the collective modes computed with RPA from the spin model is further confirmed as follows: we  fitted the  lowest energy branches to the dispersion energy predicted by linear spin wave theory for the  FM Heisenberg model. In that model, the second branch would be symmetric to the first one, given the isomorphism with the electron-hole symmetric bands of graphene. However, in Fig. \ref{fig:magnondispersions} it is apparent that the second branch departs from this picture. 

In conclusion, we have shown that the collective properties  of structures made with open-shell nanographenes with orbital degeneracy have a physical behaviour richer than their non-orbitally degenerate counterparts.  We have focused on the case of nitrogen doped triangulenes, a recently synthesized molecule \cite{wang2022,lawrence23, calupitan2023emergence}, as a building block for two types of structures, dimers and honeycomb lattices. Our calculations show that antiferromagnetic states cannot be obtained by flipping over the spins of ferromagnetic solutions in one sublattice, at odds with the case of conventional magnetic nanographenes. Both our  multiconfigurational and RPA calculations show  that the number of low energy degrees of freedom is increased, compared to the naive spin model. Altogether, our results  provide strong evidence that in open-shell nanographenes with orbital degeneracy,  spin degrees of freedom are strongly coupled to an orbital pseudospin, invalidating the use of typical spin models. Importantly, the spin-orbital interplay, reflected for example in the spin-dependence of the charge maps of 3AT dimers, provides a built-in mechanism for spin-to-charge conversion that would ease the readout of their spin state, using for instance a nearby charge detector \cite{buks1998dephasing}, with potential for applications in quantum technologies \cite{elzerman2004single}.

{\em Note added}. During the last stages of this work, we became aware of a work\cite{yu23} with  results in agreement with our DFT calculations for the 3AT crystal.

\begin{acknowledgments}
J.F.-R., J.C.G.H. and A.T.C.  acknowledge financial support from 
 FCT (Grant No. PTDC/FIS-MAC/2045/2021),
SNF Sinergia (Grant Pimag,  CRSII5\_205987)
 the European Union (Grant FUNLAYERS
- 101079184).
J.F.-R. acknowledges financial funding from 
Generalitat Valenciana (Prometeo2021/017
and MFA/2022/045)
and
MICIN-Spain (Grants No. PID2019-109539GB-C41 
A.M.-S. acknowledges financial support by Ram\'on y Cajal programme (grant RYC2018-024024-I; MINECO, Spain), Agencia Estatal de Investigación (AEI), through the project PID2020-112507GB-I00 (Novel quantum states in heterostructures of 2D materials), and Generalitat Valenciana, program SEJIGENT (reference 2021/034), project Magnons in magnetic 2D materials for a novel electronics (2D MAGNONICS)and by Generalitat Valenciana, project SPINO2D, reference MFA/2022/009.
J.F.-R. and A M.-S. acknowledge funding from the 
the Advanced Materials programme  supported by MCIN with funding from European Union NextGenerationEU (PRTR-C17.I1) and by Generalitat Valenciana (MFA/2022/045).
%
%
DJ acknowledges financial support by Grant No. PID2020-112811GB-I00 funded
by MCIN/AEI/10.13039/501100011033 and by Grant No. IT1453-22 from the Basque Government.
G.C. acknowledges financial support by the Werner Siemens Foundation (project CarboQuant).
\end{acknowledgments}

\bibliographystyle{apsrev4-2}
\bibliography{bibshort}

\end{document}